\documentclass[12pt,preprint]{aastex}

\shorttitle{Initial Conditions for GC Formation}
\shortauthors{Mashchenko and Sills}

\begin{document}

\title{The Universality of Initial Conditions for Globular Cluster Formation}

\author{Sergey Mashchenko\altaffilmark{1} and Alison Sills\altaffilmark{1}}

\altaffiltext{1}{Department of Physics and Astronomy, McMaster University,
Hamilton, ON, L8S 4M1, Canada; syam,asills@physics.mcmaster.ca}

\begin{abstract}
We investigate a simple model for globular cluster (GC) formation. We simulate the
violent relaxation of initially homogeneous isothermal stellar spheres and show
that it leads to the formation of clusters with radial density profiles that
match the observed profiles of GCs. The best match is achieved for
dynamically unevolved clusters.  In this model, all the observed correlations
between global GC parameters are accurately reproduced if
one assumes that all the clusters initially had the same value of the stellar
density and the velocity dispersion. This suggests that the gas which formed
GCs had the same values of density and temperature throughout the
universe.
\end{abstract}
 

\section{INTRODUCTION AND MOTIVATION}

Globular clusters (GCs) are massive, dense clusters of stars, and in many galaxies
they are the oldest datable objects. They contain important clues to early star
formation in the universe, and to the formation history of our Milky Way and
other galaxies. The dynamical evolution of GCs has been studied
for many decades, and is considered to be well-understood. However, the
formation of GCs is still an open question.

The internal structure and dynamics of Galactic GCs are well
described by a family of lowered isothermal models (so-called King models,
\citealt{K66}). The radial profiles of King models have a flat core (characterized
by a core radius $r_0$) and a sharp cut-off at a tidal radius $r_{\rm t}$.  GCs
are presently in dynamical equilibrium. Initially they could have formed in some
non-equilibrium configuration which relaxed to a King model through violent
relaxation
\citep{MH97} followed by a slow evolution in the galactic tidal
field. 

Currently no models of GC formation explain all the observed
properties of these star clusters
\citep{DM94,M00}. Correlations between parameters which are not
significantly affected by the dynamical evolution of a cluster (such as half-mass
radius, central velocity dispersion and binding energy) are thought to reflect
the initial conditions for GC formation. Therefore, any model of
cluster formation should aim to address the observed correlations.

In this paper, we investigate the simplest possible model for GC
formation: the dynamical evolution of a homogeneous isothermal stellar
sphere. Simulations of the collapse of homogeneous stellar sphere with constant
velocity dispersion were performed twenty years ago \citep{vA82} with the
intention of reproducing the $r^{1/4}$ density distribution of elliptical
galaxies. Those efforts were not very successful --- the collapse produced
instead a core-halo structure, somewhat like a GC. In this paper,
we compare a family of such models with the observed density profiles of
GCs, and attempt to use the family of such models to reproduce
correlations between parameters of GCs in the Milky Way. 

\section {MODEL}

We propose a homogeneous isothermal stellar sphere as an initial non-equilibrium
configuration for GCs. To test this idea, we evolve stellar
spheres with a total mass $M$, an initial density $\rho_{\rm ini}$ and an
initial velocity dispersion $\sigma_{\rm ini}$ for $t_{\rm end}=10-5000$ initial
crossing times $t_{\rm cross} = (R_{\rm ini}^3 / M)^{1/2}$, where $R_{\rm ini} =
(3 M / 4 \pi \rho_{\rm ini})^{1/3}$ is the initial radius of the system.  (In
this Letter we assume a system of units where the gravitational constant $G =
1$.)  We use the parallel N-body tree-code GADGET \citep{SYW01} to run the
simulations.  The stars in the cluster are represented by $N = 10^5 - 10^6$
equal mass particles. The velocities of the particles have a Maxwellian
distribution.  The gravitational potential is softened with a softening length
of $\epsilon = 0.77 R_{\rm h} N^{-1/3}$, where $R_{\rm h}$ is the initial
half-mass radius of the cluster. (Gravity is softened to minimize numerical
noise due to two-body interactions between the particles.) The individual
timesteps are equal to $\sqrt{2\eta \epsilon /a}$, where $a$ is the acceleration
of a particle, and the parameter $\eta$ is made small enough to ensure the total
energy conservation to better than 1\%.  This configuration is the simplest
realization of possible initial conditions for GCs, and yet
results in remarkably good agreement with current cluster parameters.

It is convenient to express the mass of a cluster $M$ in units of the
virial mass $M_{\rm vir}: M = M_{\rm vir} 10^\beta$, where $\beta$ is
a mass parameter. For a homogeneous sphere $M_{\rm vir} = \sigma_{\rm
ini}^3 (4 \pi \rho_{\rm ini} / 375)^{-1/2}$. The total
energy of the system becomes positive (and the system becomes formally unbound)
for $\beta < -0.45$.

We ran a set of 17 different models (see Table~\ref{table1}) with the same
values of $\rho_{\rm ini}$ and $\sigma_{\rm ini}$, and parameter $\beta$ ranging
from --0.8 to 2. (The corresponding initial virial parameter $\nu \equiv 2K/W =
10^{-2\beta/3}$ is $\nu$ = 3.4 to 0.046, where $K$ and $W$ are initial kinetic
and potential energy.)  Models with $\beta$ = --0.8 are completely unbound
throughout the simulation, whereas models with $\beta$ = (--0.7, --0.6, --0.5)
form a bound cluster, containing less than 100\% of the total mass, after the
initial expansion phase. Models with $\beta \geq 0$ initially collapse to a
smaller half-mass radius $r_{\rm min}$ (see Table~\ref{table1}), and then
bounce. All models with $\beta \geq -0.7$ experience phase mixing and/or violent
relaxation, and within 10 -- 100 crossing times reach an equilibrium state with
a flat core and sharply declining density at large radii. The projected surface
density profiles of equilibrium clusters closely resemble the profiles of
Galactic GCs. The match is the best for the GCs
which have experienced little dynamical evolution (see Figure \ref{surface
density}).

Our models fit the observed surface density profiles of GCs
very well, despite the lack of a tidal radius as present in
King models. We do not include an external tidal field in our
simulations, so this lack is expected. The tidal cutoff in surface
density is observed in very few GCs \citep{TKD95}, due to the
contamination with background stars at the outskirts of the clusters. 

As we will show in Section~\ref{sec3}, all real GCs should have experienced an
adiabatic collapse (when the orbital angular momentum is conserved for
individual stars). To make sure that our models are in the same collapse regime
as the real clusters, all our models should satisfy the following adiabaticity
criterion \citep*{alp88}:

\begin{equation}
\sigma_{\rm ini} \gtrsim N^{-1/6} (M/R_{\rm ini})^{1/2}.
\label{ALP}
\end{equation}

\noindent In our case, this
condition can be expressed as $\beta \lesssim (1/2)\log N - (3/2)\log 5$.  The
adiabatic collapse criterion is then $\beta\lesssim 1.5$ for $N=10^5$ and
$\beta\lesssim 2.0$ for $N=10^6$. According to this criterion, all our models
undergo an adiabatic collapse (see Table~\ref{table1}).  In our models, the
collapse factor $C \equiv R_{\rm h}/r_{\rm min}$ correlates well with the virial
parameter $\nu$: $C\propto \nu^{-\gamma}$. The value of the exponent 
$\gamma =0.95\pm 0.02$  is very close to the adiabatic value of $\gamma=1$
\citep{alp88}.

From our set of models, we obtained the following correlations between the
projected half-mass radius $r_{\rm h}$, projected central velocity dispersion
$\sigma_0$, central surface density $\Sigma_0$, central density
$\rho_{\rm c}$, King core radius $r_0$, binding energy $E_{\rm b}$, and mass $M$
(the uncertainties in the exponents are $\lesssim 0.01$):

\begin{eqnarray}
r_{\rm h}  \sim {\rm constant},\quad & \sigma_0 \propto M^{0.57}, & \quad\Sigma_0 \propto M^{1.39}, \nonumber\\
\rho_{\rm c} \propto  M^{1.68},\quad & r_0 \propto M^{-0.27}, & \quad E_{\rm b} \propto M^{1.95},\label{model}\\
&E_{\rm b} \propto \sigma_0^{3.55}.& \nonumber
\end{eqnarray}

\noindent These correlations are valid for $\beta \geq 0$, with the last relation
being valid for $\beta \geq -0.6$. Most of correlations become non-linear in
log-log space for $\beta < 0$ (corresponding to $\nu > 1$).  Although for our
coldest models, $r_{\rm h}$ scales (as expected) as the initial radius (and
hence as $M^{1/3}$), the total variation in $r_{\rm h}$ over two orders of
magnitude in system mass ($\beta=0-2$) is only $\sim 0.12$~dex (see
Table~\ref{table1}; please note that $R_{\rm ini}\propto M^{1/3}$).  

Another very interesting correlation is $r_0\simeq r_{\rm
min}$ within the measurement errors for $\beta\geq 0.2$ (Table~\ref{table1}).
The above correlation can be understood if we rewrite the theoretical result $C
\equiv R_{\rm h}/r_{\rm min}\propto \nu^{-1}$ \citep{alp88} for the case of
constant $\rho_{\rm ini}$ and $\sigma_{\rm ini}$: $r_{\rm min}\propto M^{-1/3}$.
The theoretical relation is very close to the model relation $r_0 \propto
M^{-0.27}$, resulting in $r_0\propto r_{\rm min}$ as observed.

We tested the numerical convergence of our results by making two additional runs
for $\beta=1.6$. In the first one, the number of particles was reduced to
$N=10^5$, with the optimal value for the softening length of $\log
(\epsilon/R_{\rm ini})=-1.88$. Despite the fact that this run had the most
poorly resolved ``crunch'' among all our models ($r_{\rm min}/\epsilon\simeq
3.6$), all the derived model parameters were identical to the original run
parameters within measurement errors.  In the second run, we again used a
reduced number of particles $N=10^5$, but the parameter $\epsilon$ was made much
smaller than the optimal value: $\log (\epsilon/R_{\rm ini})=-2.67$. In this
run, the crunch is resolved very well --- $r_{\rm min}/\epsilon\simeq 23$, but
again we did not see significant deviations from the original run
parameters. The results of the additional runs suggest that the number of
particles $N$ and the value of the softening length $\epsilon$ we use in our
runs are adequate for resolving the violent relaxation process.

\section{COMPARISON WITH OBSERVATIONS}
\label{sec3}

We assume that the mass-to-light ratio $M/L$ is a constant, which is
well-established for observed GCs ($\log [M/L] = 0.16 \pm 0.03$,
\citealt{M00}). Then we derive the same correlations for observed
clusters as in equations~(\ref{model}) using the online version of the Milky Way
GC catalogue of W. E. Harris \citep{H96}. We used data for 109
non-core-collapsed GCs. Forty-five of these clusters have a
measured $\sigma_0$. Binding energies were calculated using equations (5a-c) of
\citet{M00}. The observed correlations are:

\begin{eqnarray}
r_{\rm h} \sim {\rm constant},\quad & \sigma_0 \propto  M^{0.43 \pm 0.05}, &\quad \Sigma_0 \propto  M^{1.31 \pm 0.10},\nonumber\\
\rho_{\rm c} \propto   M^{1.53 \pm 0.17},\quad & r_0 \propto  M^{- 0.28 \pm 0.07}, &\quad E_{\rm b} \propto  M^{2.03 \pm 0.08},\label{observed}\\
&E_{\rm b} \propto  \sigma_0^{3.61 \pm 0.15}.&\nonumber 
\end{eqnarray}

\noindent The theoretical exponents differ from the observational values by one
sigma or less (with the exception of the $\sigma_0[M]$ correlation,
where the difference is 3 sigma).

The closeness of the model correlations (equations [\ref{model}]) to the
observed correlations (equations [\ref{observed}]) can be understood if we
assume that in their initial non-equilibrium configuration, all
Galactic GCs had the same values of stellar density
$\rho_{\rm ini}$ and velocity dispersion $\sigma_{\rm ini}$.

To estimate the values of the universal parameters $\rho_{\rm ini}$
and $\sigma_{\rm ini}$, one can use in principle any two or more pairs
of model/observational correlations from equations (\ref{model}) and
(\ref{observed}). However, one has to keep in mind that few GC
parameters are well suited for comparing the model and
observed correlations. Most Galactic GCs are in advanced
stages of their dynamical evolution. As GCs evolve, some
of their parameters deviate from initial equilibrium values. This
process affects mainly the central core region of a cluster where
encounters between stars are the most frequent. Analytical and
numerical calculations show that at some point a GC
should experience a runaway core collapse due to gravothermal
instability \citep{S87}. Around 20\% of Galactic GCs are
believed to be in a post-core-collapse state \citep{H96}.  The
analytical theory of core collapse \citep{S87} can be used to find
GC parameters which are least affected by dynamical
evolution. We obtained the following relations:

\begin{eqnarray}
\rho_{\rm c} \propto  r_0^{- 2.21},\quad & \Sigma_0 \propto  r_0^{- 1.21}, & \quad \sigma_0 \propto  r_0^{- 0.10}.\label{core}
\end{eqnarray}

\noindent To obtain an analogous relation for binding energy $E_{\rm b}$, we used
equations (5a-c) of \citet{M00}. Assuming that during core collapse the
tidal radius $r_{\rm t}$ does not change, we derived

\begin{equation}
E_{\rm b} \propto  r_0^{- 0.05},
\label{core2}
\end{equation}

\noindent which is accurate for the range of $r_{\rm t} / r_0 = 10 - 100$
corresponding to 75\% of Galactic GCs. Equations
(\ref{core}) and (\ref{core2}) show that the parameters $E_{\rm b}$ and
$\sigma_0$ are least affected by dynamical evolution. Numerical
simulations of core collapse show that the half-mass radius $r_{\rm h}$ is
also a very slowly evolving parameter \citep{S87}.

We used two following correlations to estimate the values of
$\rho_{\rm ini}$ and $\sigma_{\rm ini}$: $E_{\rm b}(\sigma_0)$ and $r_{\rm h} =$
constant. The $\chi^2$ fitting gave the following estimates: $\log
\rho_{\rm ini} = 1.14 \pm 0.26$, and $\log \sigma_{\rm ini} = 0.28 \pm
0.11.$ (The units for $\rho_{\rm ini}$ and $\sigma_{\rm ini}$ are M$_{\odot}$
pc$^{-3}$ and km s$^{-1}$.)  The relation between the masses of real GCs and the
model mass parameter $\beta$ is then $M\simeq 3.5\times
10^{\beta+4}$~M$_{\odot}$.  The corresponding minimum initial cluster mass
resulting in a bound cluster (the model with $\beta = -0.7$) is $M_{\rm min}
\sim 6,900$ M$_{\odot}$ , with a one-sigma interval of 3,000 -- 16,000
M$_{\odot}$. The criterion of an adiabatic collapse of \citet{alp88}
(equation~[\ref{ALP}]) can be reexpressed as $M\lesssim 3\sigma_{\rm
ini}^6/(4\pi \rho_{\rm ini}m_1)$, where $m_1$ is a typical stellar mass in the
cluster. For our values of $\rho_{\rm ini}$ and $\sigma_{\rm ini}$, a collapse
is adiabatic if $M\lesssim 1.6\times 10^7$~M$_\odot$ (we assumed
$m_1=0.6$~M$_\odot$).  As one can see, our adiabatic collapse models are
adequate for the whole range of GC masses.

The simplest physical interpretation of the universal initial density
value is to assume that the major burst of star formation in a
contracting proto-GC molecular cloud occurs when the gas
density reaches a certain critical value $\rho_{\rm cr} > \rho_{\rm
ini}$ (the non-equality sign is to account for less than a 100\%
efficient star formation and mass loss due to stellar evolution). The
initial velocity dispersion $\sigma_{\rm ini}$ can be assumed to be
commensurable with the sound speed in the turbulent
proto-GC cloud, and hence with its temperature $T$, at the
moment when the star burst occurs. For a purely molecular gas of
primordial composition (Y$_{\rm p}$ = 0.291), we obtained the following
estimates of the universal gas density $\rho_{\rm cr}$ and temperature
$T$ for star-forming gas in proto-GCs: $\rho_{\rm cr} >
230$ cm$^{-3}$, and $T \sim $ 1,000 K. The corresponding gas pressure is
$P > 2.3 \times 10^5$ K cm$^{-3}$.

\begin{figure}
\plotone{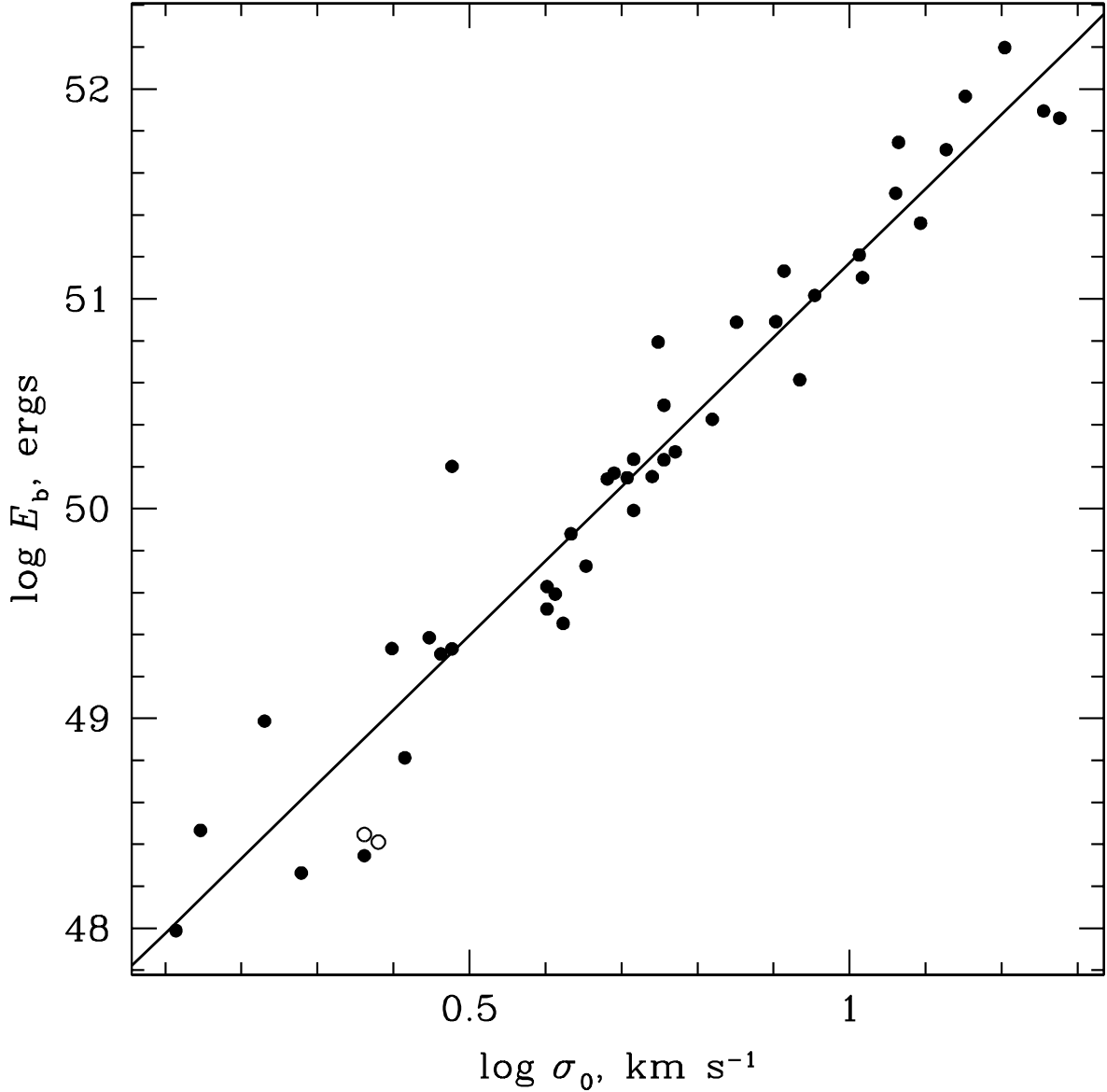}
\caption{Cluster binding energy as a function of velocity dispersion
for 45 Galactic GCs with a measured $\sigma_0$ (circles) and our model
(solid line). The open circles are the most dynamically evolved
clusters with $\log t_{\rm c} < 8.3$, where $t_{\rm c}$ is the core relaxation time in
years. \label{binding velocity}}
\end{figure}

\begin{figure}
\plotone{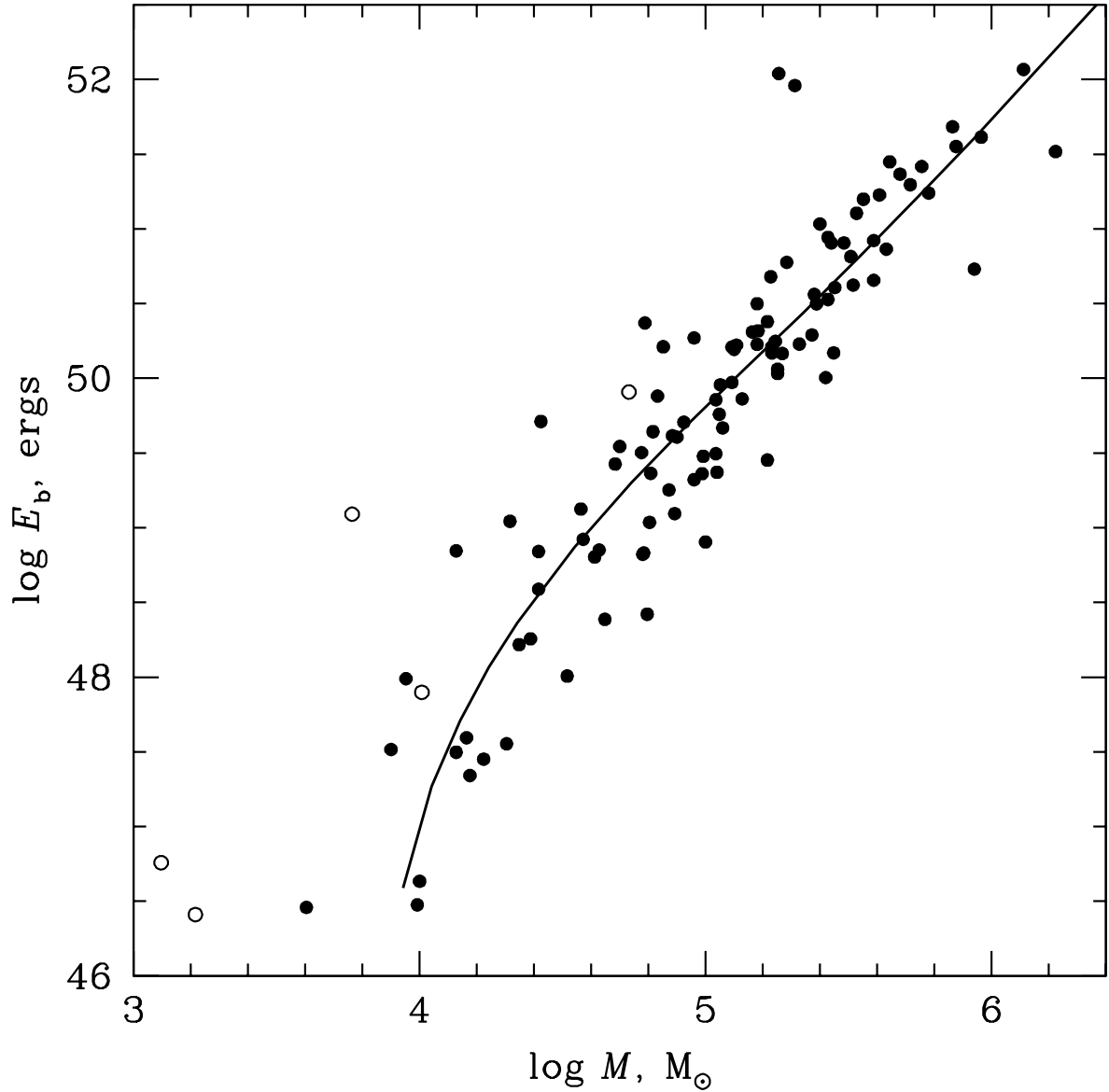}
\caption{Cluster binding energy as a function of total cluster mass
for 109 non-core-collapsed GCs (circles) and our model (solid
line). The open circles are those clusters with $\log t_{\rm c} < 8.3$ (as in
Figure 2). \label{binding mass}}
\end{figure}

\begin{figure}
\plotone{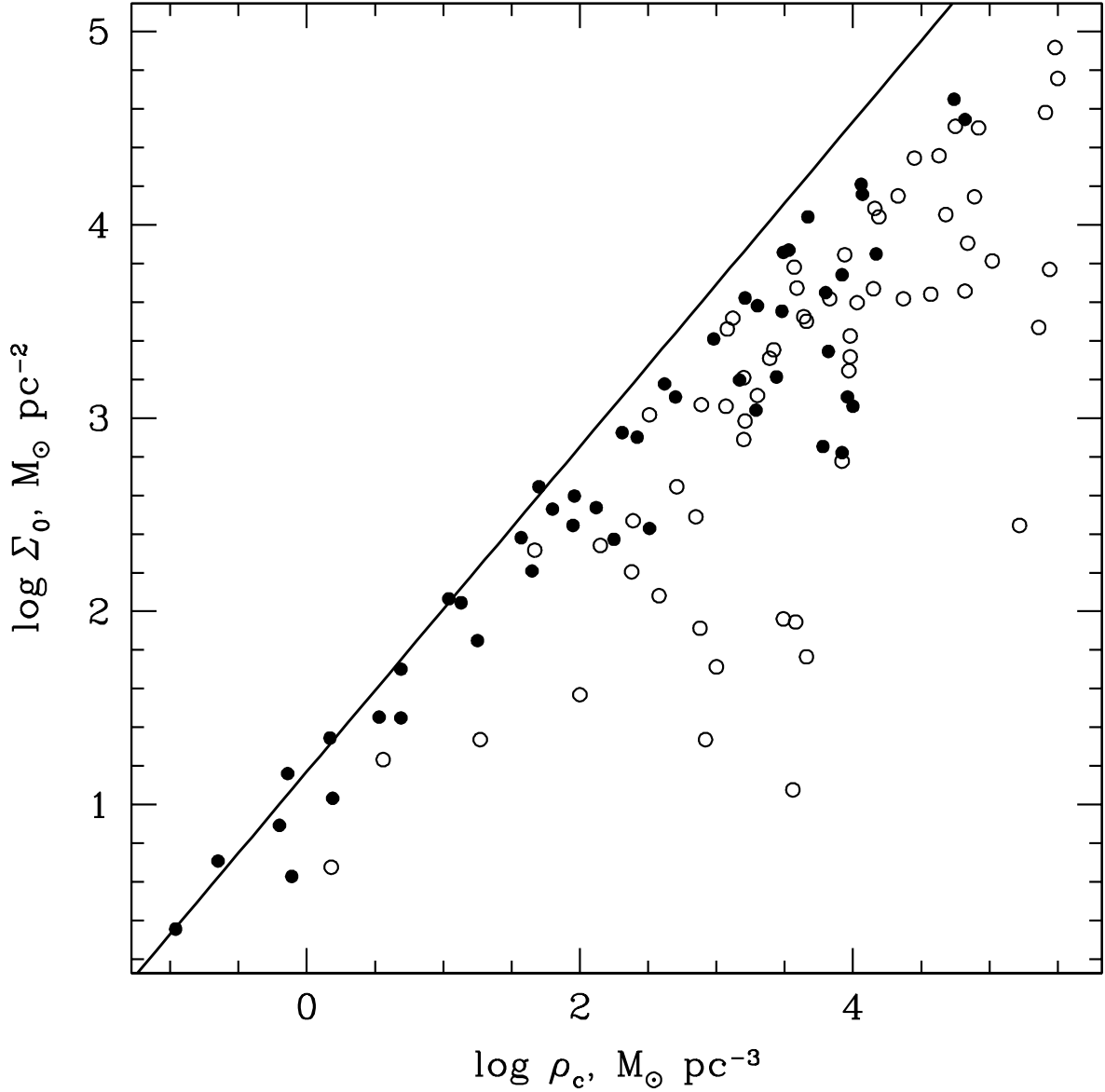}
\caption{Central surface density as a function of central density
for 109 non-core-collapsed  GCs (circles) and our model (solid
line). The open circles are more dynamically evolved clusters with $\log
t_{\rm c} < 9.2$. \label{bright density}}
\end{figure}

In Figures \ref{binding velocity}--\ref{bright density} we compare three
different observed correlations for Galactic GCs with the model
predictions rescaled to $\log \rho_{\rm ini} = 1.14$ and $\log \sigma_{\rm ini}
= 0.28$: $E_{\rm b}(\sigma_0), E_{\rm b}(M)$ and $\Sigma_0(\rho_{\rm c})$. In
Figures \ref{binding velocity} and \ref{binding mass} (showing the correlations
between the least evolved cluster parameters) the agreement between the model
and data is excellent, with a low statistical significance suggestion that the
most evolved clusters, shown as open circles, tend to deviate from the model
$E_{\rm b}(M)$ relation (Figure \ref{binding mass}). Figure \ref{bright density}
shows a correlation between the most evolved cluster parameters --- central
surface density $\Sigma_0$ and central density $\rho_{\rm c}$ (see equations
[\ref{core}]). The situation with Figure \ref{bright density} is in accord with
our model predictions: all clusters deviate from the model, "zero-age", relation
in a systematic fashion, with more evolved clusters located farther from the
model line. (Please note that the correlations from Figures \ref{binding mass}
and \ref{bright density} were not used to fit the model to the data.)

\section{SUMMARY}

We have shown that a simple model of violent relaxation of a constant density,
constant velocity dispersion stellar sphere produces clusters which are
remarkably similar to the observed Galactic GCs. They have the correct surface
density profiles, and a large number of correlations between cluster parameters
are accurately reproduced by our models as long as we assume a constant initial
density and initial velocity dispersion.  This result implies that all GCs
were formed from gas with a universal value of density and
temperature. In reality, the constant density, constant velocity dispersion
initial setup of our models may correspond to isothermal turbulent cores of
giant molecular clouds with a Gaussian density profile. Our simple picture of
GC formation suggests that the conditions in the cluster-forming
gas must be quite uniform across a large portion of the early universe.

\acknowledgements We thank W. E. Harris, R. E. Pudritz, and
D. E. McLaughlin for useful discussions, and the anonymous referee for
improvements to the paper. S. M. is partially supported by SHARCNet. The
simulations reported in this paper were performed at CITA.

\begin{table}
\caption{Model parameters\label{table1}} 
\begin{tabular}{rcccccrrrr}
\tableline
\multicolumn{5}{c|}{Input parameters} & \multicolumn{5}{c}{Derived parameters}\\
\tableline
$\beta$&   $N$   & $\nu$ & $\frac{t_{\rm end}}{t_{\rm cross}}$ & $\log\frac{\epsilon}{R_{\rm ini}}$ & $\log\frac{r_{\rm min}}{R_{\rm ini}}$ & $\log\frac{r_0}{R_{\rm ini}}$ & $\log\frac{r_{\rm h}}{R_{\rm ini}}$ & $\log\frac{\sigma_0}{\sigma_{\rm ini}}$ & \rule[-6pt]{0pt}{15pt} $\log\frac{\rho_{\rm c}}{\rho_{\rm ini}}$\\
\tableline
--0.8   &  $10^5$ & 3.4  & 5000   &  --1.88  &{\nodata} &{\nodata}  & {\nodata}&{\nodata}  &{\nodata}\\
--0.7   &  $10^5$ & 2.9  & 1000   &  --1.88  &{\nodata} &     0.50  & {\nodata}&   --1.10  &   --2.9  \\
--0.6   &  $10^5$ & 2.5  & 1000   &  --1.88  &{\nodata} &     0.21  &  1.09 &   --0.58  &   --1.3  \\
--0.5   &  $10^5$ & 2.2  & 1000   &  --1.88  &{\nodata} &     0.05  &  0.34 &   --0.37  &   --0.7   \\
--0.4   &  $10^5$ & 1.8  & 1000   &  --1.88  &{\nodata} &   --0.06  &  0.08 &   --0.25  &   --0.3   \\
--0.3   &  $10^5$ & 1.6  & 1000   &  --1.88  &{\nodata} &   --0.12  &--0.06 &   --0.15  &     0.0   \\
--0.2   &  $10^5$ & 1.4  & 1000   &  --1.88  &{\nodata} &   --0.20  &--0.15 &   --0.07  &     0.2   \\
 0.0    &  $10^5$ & 1.0  &  200   &  --1.88  & --0.25   &   --0.31  &--0.29 &     0.07  &     0.6   \\
 0.2    &  $10^5$ & 0.74 &  200   &  --1.88  & --0.42   &   --0.42  &--0.40 &     0.19  &     0.9   \\
 0.4    &  $10^5$ & 0.54 &  200   &  --1.88  & --0.58   &   --0.57  &--0.49 &     0.30  &     1.3  \\
 0.6    &  $10^5$ & 0.40 &  200   &  --1.88  & --0.71   &   --0.70  &--0.57 &     0.41  &     1.6 \\
 0.8    &  $10^5$ & 0.29 &   50   &  --1.88  & --0.84   &   --0.79  &--0.64 &     0.52  &     1.9  \\
 1.0    &  $10^5$ & 0.22 &   50   &  --1.88  & --0.96   &   --0.92  &--0.72 &     0.63  &     2.3  \\
 1.2    &  $10^5$ & 0.16 &   50   &  --1.88  & --1.08   &   --1.03  &--0.79 &     0.74  &     2.6  \\
 1.4    &  $10^5$ & 0.12 &   50   &  --1.88  & --1.19   &   --1.16  &--0.84 &     0.85  &     2.9  \\
 1.6    &  $10^6$ & 0.086&   15   &  --2.21  & --1.32   &   --1.28  &--0.88 &     0.98  &     3.3 \\
 2.0    &  $10^6$ & 0.046&   10   &  --2.21  & --1.54   &   --1.53  &--0.94 &     1.20  &     4.0  \\
\tableline
\end{tabular}
\tablecomments{Here $r_0$, $\sigma_0$, and $\rho_{\rm c}$ are the core parameters for relaxed models: King core radius,
projected central velocity dispersion, and central density, respectively; $r_{\rm min}$ is the minimum half-mass radius
during the collapse, and $r_{\rm h}$ is the projected half-mass radius for relaxed models.}
\end{table}

\newpage

\begin{figure}
\plotone{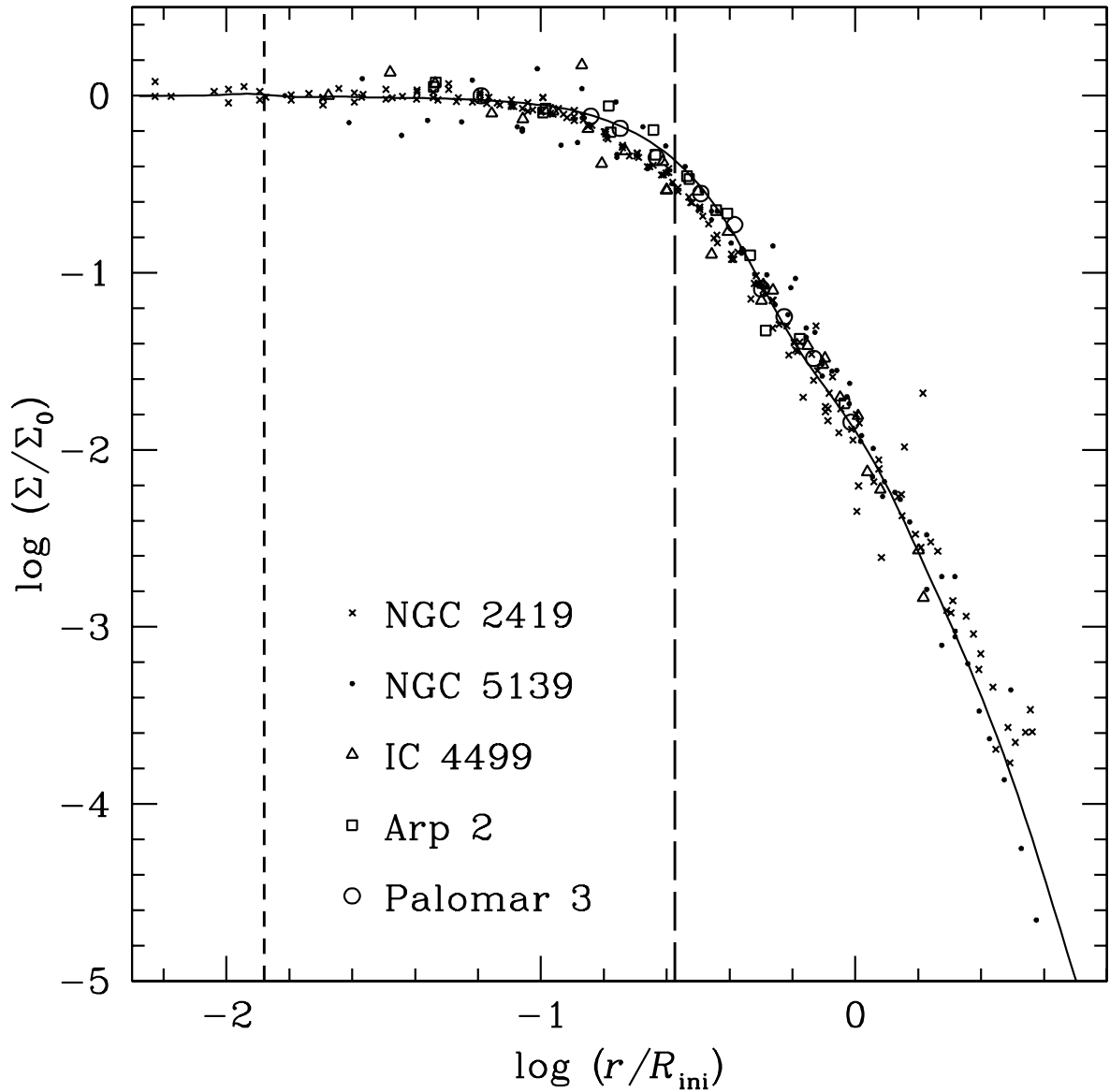}
\caption {Surface density profile (solid line) for a model
with $\beta$ = 0.4. This model matches well the surface density profiles of
five GCs (points, \citealt*{TKD95}) which are among the least
dynamically evolved ones.  Surface density is normalized to 1 at the
center of the cluster. The vertical short-dashed and long-dashed lines show the
values of the softening length $\epsilon$ and the model core radius $r_0$,
respectively. \label{surface density} }
\end{figure}

\end{document}